\documentclass[preprint,preprintnumbers,amsmath,amssymb,aps]{revtex4-2}
\usepackage{graphicx}
\usepackage{dcolumn}
\usepackage{bm}

\begin{document}

\title{Modified Dirac fermions in the crystalline xenon and graphene Moir\'{e} heterostructure}

\author{Hayoon Im$^{1,{\dagger}}$}
\author{Suji Im$^{1,{\dagger}}$}
\author{Kyoo Kim$^{2,{\dagger}}$}
\author{Ji-Eun Lee$^{3}$}
\author{Jinwoong Hwang$^{4}$}
\author{Sung-Kwan Mo$^{3}$}
\author{Choongyu Hwang$^{1,5,*}$}

\medskip

\affiliation{$^1$ Department of Physics, Pusan National University, Busan, Korea}
\affiliation{$^2$ Korea Atomic Energy Research Institute, Daejeon, Korea}
\affiliation{$^3$ Advanced Light Source, Lawrence Berkeley National Laboratory, Berkeley, CA, USA}
\affiliation{$^4$ Department of Physics, Kangwon National University, Chuncheon, Korea}
\affiliation{$^5$ Quantum Matter Core-Facility, Busan, Korea}
\affiliation{$^{\dagger}$ These authors contributed equally to this work.}


\begin{abstract}
The interface between two-dimensional (2D) crystals often forms a Moir\'{e} superstructure that imposes a new periodicity, which is a key element in realizing complex electronic phases as evidenced in twisted bilayer graphene. A combined angle-resolved photoemission spectroscopy measurements and first-principles calculations reveal the formation of a Moir\'{e} superstructure between a 2D Dirac semi-metallic crystal, graphene, and a 2D insulating crystal of noble gas, xenon. Incommensurate diffraction pattern and folded Dirac cones around the Brillouin zone center imply the formation of hexagonal crystalline array of xenon atoms. The velocity of Dirac fermions increases upon the formation of the 2D xenon crystal on top of graphene due to the enhanced dielectric screening by the xenon over-layer. These findings not only provide a novel method to produce a Moir\'{e} superstructure from the adsorption of noble gas on 2D materials, but also to control the physical properties of graphene by the formation of a graphene-noble gas interface.
\end{abstract}

\maketitle

\section{Introduction}

The interface between two different materials invites exotic phenomena that do not exist in each material alone. For example, the interface between two different oxides can host superconductivity~\cite{Reyren,Gozar,Caviglia} and magnetism~\cite{Brinkman} in the presence of electronic phase separation~\cite{Ariando}. Most of the rich physics at the oxide interfaces are closely related to strain that drastically modify the electronic and magnetic properties at the interface~\cite{Chakhalian}. Recent efforts in isolating two-dimensional (2D) materials, on the other hand, have allowed to stack 2D crystals in the absence of strains~\cite{Batzill, Jin, Pierucci, Wilson}. Instead, their 2D nature makes each crystal very sensitive to the potential induced by an adjacent layer, so that the misorientation between stacked 2D crystals imposes a new periodicity~\cite{Graham, Wang, Lu}. Resultant Moir\'{e} potential with a long periodicity folds the electron band structure of each crystal into a much smaller Brillouin zone, resulting in heavily modified electronic structures and a variety of novel electronic phases, including unconventional superconductivity~\cite{Cao1}, Mott insulating phase~\cite{Cao2}, ferromagnetism~\cite{Sharpe}, and ferroelectricity~\cite{Weston}.

Periodic adsorption of foreign atoms on a 2D material plays a similar role as the stacked 2D crystals~\cite{Sugawara, Shin, Bao, Qu}. Especially, the van der Waals nature of 2D crystals made it possible to impose additional periodic potential without modifying their chemical environment except chemical potential shift by the change of work function of the whole system. For example, lithium atoms on single- and double-layer graphene give a short range periodic potential, inducing a $(\sqrt{3} \times \sqrt{3}) R30^{\circ}$ phase~\cite{Sugawara}. Such additional potential promotes Kekul\'{e}-type charge modulation in underlying graphene, that has been evidenced by energy gap opening at the Dirac energy of the folded Dirac cones at the Brillouin zone center~\cite{Bao,Qu}.

The periodic modulation of potential can also occur when noble gases are adsorbed on the surface of 2D materials. While early studies focussed mostly on the solidification or freezing that constructs a face-centered cubic structure of noble gases~\cite{Eatwell, Sears, Kaindl, Meixner}, xenon adsorbed graphite~\cite{Bracco,Hamichi} exhibits an unusual spectral features with the inclusion of additional metallic atoms~\cite{Patthey}. Such features have been attributed to the energy gap opening at the Dirac energy by the charge transfer from the metal atoms~\cite{Pivetta}. Interestingly, xenon atoms adsorbed on graphene were theoretically predicted to exhibit two-dimensional incommensurate solid phase with the triplet point estimated to be $\sim$100~K at a pressure of 7.6$\times$10$^{-3}$~Torr~\cite{Maiga}.

Here we combine two complimentary tools, angle-resolved photoemission spectroscopy (ARPES) and first-principles calculations, to elucidate the low-temperature electronic structure of the graphene-xenon Moir\'{e} heterostructure. Xenon adsorption on graphene leads to an incommensurate superstructure peaks in the low-energy-electron diffraction (LEED) pattern and folded Dirac cones around the Brillouin zone center in the electron band structure. Such spectroscopic features indicate the formation of a Moir\'{e} superstructure by the presence of the 2D crystalline array of xenon atoms. Concomitantly, the slope of the Dirac cone becomes steeper, demonstrating enhanced dielectric screening by the 2D xenon crystal that results in lighter effective mass of quasiparticles in graphene within the Fermi liquid theory. The formation of 2D xenon crystal is theoretically supported by the dispersive electron band structure of xenon over-layer. These results provide a simple but straightforward way to tune the physical properties of a 2D material via the Moir\'{e} superstructure induced by the crystalline noble gases.

\section{Methods}
\subsection{ARPES measurements}
Graphene samples were prepared by the epitaxial growth method on an SiC(0001) substrate~\cite{Xiaozhu}. Graphene was exposed to xenon gas under a pressure of 1$\times$10$^{-7}$~Torr at 40~K. Xenon exposure of 6 Langmuir (L) did not show notable change in the LEED pattern indicating that xenon atoms were not condensed yet, which possibly suggests that the xenon coverage on graphene is far less than 1 monolayer (ML). Additional 12~L exposure of xenon gas led to the LEED pattern and ARPES data corresponding to the Moir\'{e} heterostructure shown in Figs.~1 and 2, indicating that xenon crystal is finally condensed. LEED patterns have been measured at 40~K and room temperature at several different energies. High-resolution ARPES experiments have been performed at 40~K using a Helium discharge lamp. Energy and momentum resolutions were 68~meV and $\sim$0.03~${\rm \AA}^{-1}$, respectively, at 40~K. All the measurements were performed in ultra-high vacuum with a base pressure of 2.4$\times$10$^{-11}$~Torr.

\medskip

\subsection{Band structure calculations}
The electronic structure of the crystalline xenon single-layer on graphene has been calculated using OpenMX code~\cite{Openmx}. The atomic position of experimentally suggested supercell is relaxed within Perdew-Burke-Ernzerhof exchange correlation potential~\cite{PBE}, and 2$\times$2$\times$1 {\it k}-point mesh has been used. The effect of the spin-orbit coupling for light atom, carbon, and filled shell atom, xenon, was ignored in the structure optimization and the van der Waals interaction within D3 scheme was taken into account. To understand the effect of graphene on a xenon atom, total energies were compared for the systems where xenon sits on the hollow, on-top, and bridge sites in a 4$\times$4 graphene supercell. The system is the most stable with the shortest graphene-xenon distance when a xenon atom sits on the hollow site, while the other two high symmetric sites, on-top and bridge sites, differ only by about 0.01~${\rm \AA}$ at most. The energy difference from other configurations ($\leq$1.29~meV) is negligible as well compared to the formation energy of the xenon lattice ($\sim$12.0~meV)~\cite{Kern}.

\section{Results and Discussion}

\begin{figure*}[b]
  \includegraphics[width=1\columnwidth]{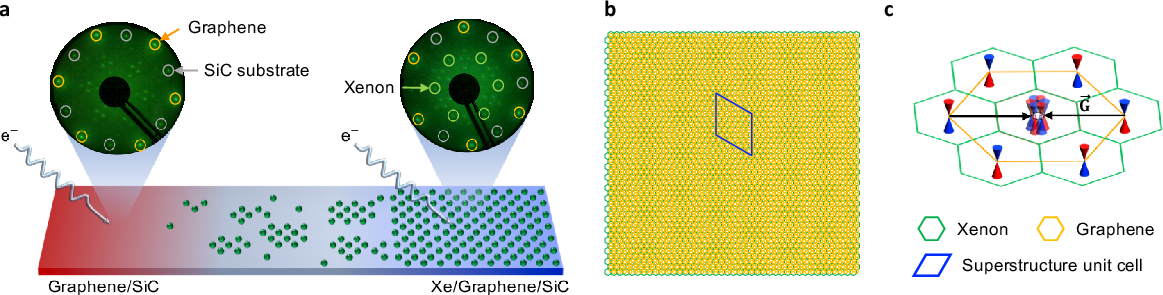}
  \caption{(a) LEED patterns of the graphene-xenon interface at 295~K (left) and 40~K (right). Orange, green, and gray circles denote LEED spots corresponding to graphene, xenon, and SiC, respectively. (b) Moir\'{e} structure simulated by the relative lattice constant of graphene and overlying xenon, extracted from the LEED pattern shown in the panel {\bf a}. Orange and green hexagons denote the real-space unit cell of graphene and overlying Xe layer, respectively. The blue lozenge is the unit cell of the Moir\'{e} superstructure. (c) Schematic drawing of the characteristic Dirac cone at the Brillouin zone corner of graphene and additional Dirac cones induced by the Umklapp process by $\vec{{\rm {\bf G}}}$, the reciprocal lattice vector of xenon over-layer.}  
  \label{fig:fig1}
\end{figure*}

Figure~1{\bf a} shows LEED patterns of single-layer graphene epitaxially grown on an SiC(0001) substrate exposed to xenon gas at a sample temperature of 40~K in ultra-high vacuum with a base pressure of 2.4$\times$10$^{-11}$~Torr. The gray and orange circles denote the characteristic LEED spots corresponding to the SiC substrate and graphene, respectively~\cite{Forbeaux}. Xenon adsorption brings about additional LEED spots denoted by green circles in the right panel that lose their intensity upon increasing temperature above $\sim$60~K, suggesting thermal desorption of xenon atoms consistent with previous reports~\cite{Bracco,Patthey}. The LEED pattern of as-grown graphene is fully recovered at 295~K as shown in the left panel. The xenon-induced LEED spots, however, do not constitute the commensurate ($\sqrt{3}$$\times$$\sqrt{3}$)R30$^{\circ}$ phase reported in the previous work~\cite{Patthey}. When noble gas is adsorbed on graphene, the center of the hexagonal unit cell, so-called hollow site, is the most favorable adsorption site~\cite{Maiga}. Considering relatively longer atomic radius of most of the adsorbates than carbon that constitutes graphene, one adatom per three~\cite{Sugawara, Bao, Qu, Yb} or four~\cite{Shin} graphene unit cells is typical of adatom-decorated graphene, resulting in the ($\sqrt{3}$$\times$$\sqrt{3}$)R30$^{\circ}$ or  2$\times$2 phases. In order for the xenon-induced additional LEED spots to be those of the ($\sqrt{3}$$\times$$\sqrt{3}$)R30$^{\circ}$ phase, the lattice constant is supposed to be 4.26~${\rm \AA}$. On the other hand, the lattice constant of the xenon-induced phase was determined to be 4.60~${\rm \AA}$ from the comparison to the inter-spot distance of graphene, of which lattice constant is 2.46~${\rm \AA}$~\cite{Wallace}. Hence the observed LEED pattern seemingly corresponds to an incommensurate phase with a hexagonal unit cell whose lattice constant is $\sim$8.0\% longer than that of the ($\sqrt{3}$$\times$$\sqrt{3}$)R30$^{\circ}$ phase.

The unit cells of graphene and xenon-induced phase are schematically drawn by orange and green hexagons, respectively, in Fig.~1{\bf b}, based on the lattice constant extracted from the LEED pattern. Interestingly, as denoted by a blue lozenge, one can find a hexagonal color pattern with a periodicity of 13 graphene unit cells. Such a periodic modulation of a solid material is typically observed when two well-defined crystals with different lattice constant or slight misorientation are overlaid to produce a Moir\'{e} superstructure. This simple cartoon suggests an interesting possibility that xenon atoms prefer to form a crystalline array instead of occupying randomly distributed hollow sites and that the resultant reciprocal lattice vector $\vec{\rm {\bf G}}$ of the xenon crystal can be used to fold the Dirac cone of graphene towards the Brillouin zone center via the Umklapp process~\cite{Ashcroft} as schematically shown in Fig.~1{\bf c}. Due to the lattice mismatch discussed above, the Dirac cone is predicted to be duplicated not exactly at the $\Gamma$ point in contrast to the case of ($\sqrt{3}$$\times$$\sqrt{3}$)R30$^{\circ}$ phase observed, for example, in lithium-adsorbed graphene~\cite{Sugawara, Bao, Qu}. 

\begin{figure*}[b]
  \includegraphics[width=1\columnwidth]{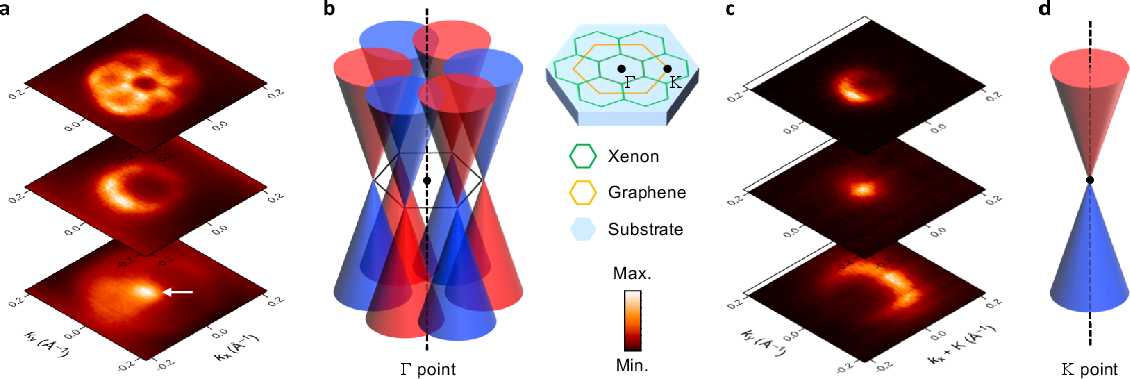}
  \caption{(a) Constant energy ARPES intensity maps of the graphene-xenon interface around the Brillouin zone center ($\Gamma$ point) of graphene taken at 40~K. (b) Schematic drawing of the six Dirac cones around the Brillouin zone center of both graphene and xenon. Inset shows the unit cell of graphene (orange) and xenon (green) in reciprocal lattice space. (c) Constant energy ARPES intensity maps of the graphene-xenon interface around the Brillouin zone corner (${\rm K}$ point) of graphene taken at 40~K. (d) Schematic drawing of the characteristic Dirac cone at the Brillouin zone corner of graphene.}
  \label{fig:fig2}
\end{figure*}

Indeed, the Fermi surface of the graphene-xenon heterostructure taken at 40~K shows six circles around the $\Gamma$ point, instead of a single circle at the $\Gamma$ point predicted for the ($\sqrt{3}$$\times$$\sqrt{3}$)R30$^{\circ}$ phase, as shown in Fig.~2{\bf a} and schematically drawn in Fig.~2{\bf b}. It should be noted that epitaxial graphene is electron-doped by the SiC substrate by the formation of a Schottky barrier~\cite{Seyller}, so that the Dirac energy, i.\,e.\,, the energy where the conduction and valence bands meet at a point, is observed below the Fermi energy, $E_{\rm F}$. Hence the Fermi surface for a single Dirac cone is expected to be a single circle. In fact, at the K point of graphene, constant energy maps taken at 40~K show an electron-doped single Dirac cone as shown in Fig.~2{\bf c} and schematically drawn in Fig.~2{\bf d}, consistent with the Fermi surface known for the single-layer graphene epitaxially grown on an SiC(0001) substrate~\cite{JW_gap}. The observation of the six Dirac cones demonstrates the formation of a two-dimensional solid crystal of xenon and the graphene-xenon Moir\'{e} heterostructure in line with the LEED result shown in Fig.~1{\bf a}.

At the Dirac energy, $E-E_{\rm F}=-0.4~{\rm eV}$, the constant energy map (Fig.~2{\bf a}) shows fuzzy six spots that are connected by finite spectral intensity, rather than six separate spots expected from the six Dirac cones, while individual energy-momentum dispersion (Fig.~3{\bf c}) clearly displays the Dirac point. Electron correlations that are typically enhanced at higher energy and hence broaden the photoemission spectra might have prevented observing clear, well-separated spots. Upon further increasing the binding energy to $E-E_{\rm F}=-1.0~{\rm eV}$, in which the six Dirac cones meet at a single point, the $\Gamma$ point, one can observe a bright spot at the overlapping point as denoted by the white arrow in Fig.~2{\bf a}, despite the circles corresponding to six Dirac cones are not clearly distinguished.

In order to understand the effect of the xenon crystal on the electronic properties of graphene, further ARPES analysis has been performed as shown in Fig.~3. The area of the Fermi surface at the K point gives information on charge carrier density in graphene by $n=2\times \frac{\pi r^2_k}{\frac{3\sqrt{3}}{2}|{\rm K}|^2}\times \frac{2}{{\frac{3\sqrt{3}}{2} b^2}}\approx r^2_k \times 3.18 \times 10^{15}~{\rm cm}^{-2}$, where $r_k$ is the radius of the circle observed at the Fermi surface, $|{\rm K}|$ is the $\Gamma$-K distance, and $b$ is the inter-carbon distance which is 1.42~${\rm \AA}$. Hence almost the same area of the Fermi surface within the experimental resolution before and after xenon adsorption implies that overlying xenon does not notably change the chemical potential of graphene either by charge transfer from/to xenon or by affecting the Schottky barrier between graphene and the substrate, different from a previous work reported the work function change of 0.2~eV~\cite{AaronXe}.

\begin{figure*}[b]
  \includegraphics[width=1\columnwidth]{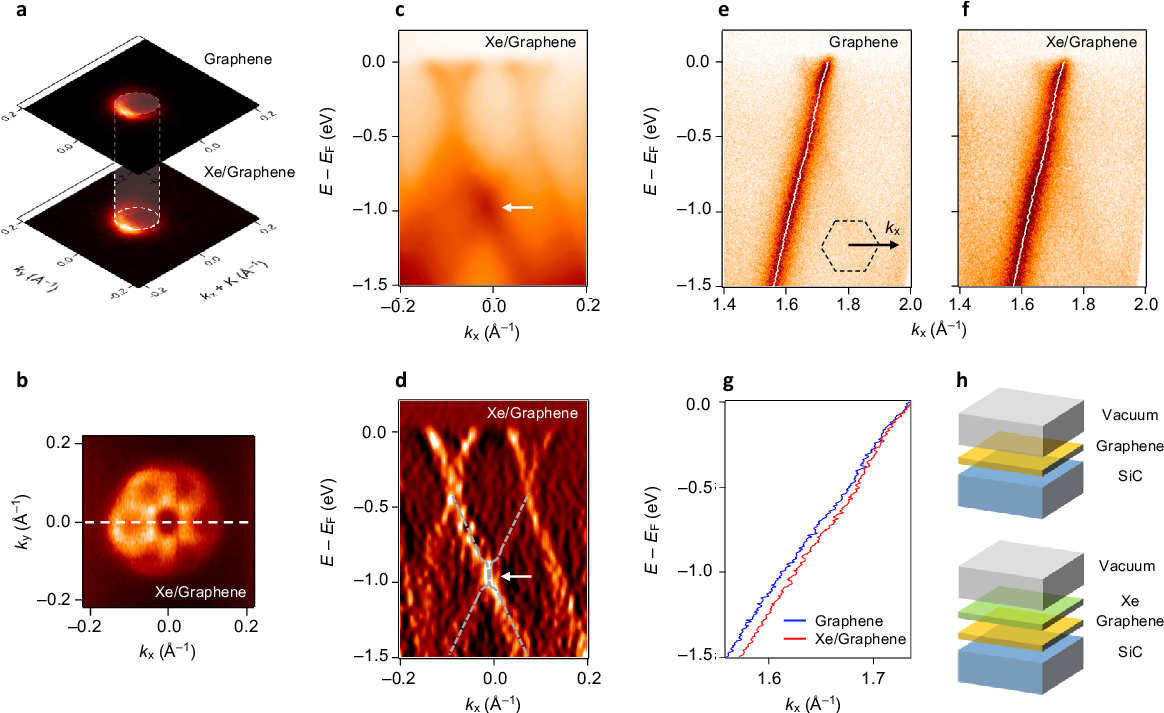}
  \caption{(a) Constant energy ARPES intensity maps taken at $E_{\rm F}$ around the ${\rm K}$ point of as-grown graphene (upper panel) and the graphene-xenon Moir\'{e} heterostructure (lower panel). (b) A constant energy ARPES intensity map  taken at $E_{\rm F}$ around the $\Gamma$ point of the graphene-xenon Moir\'{e} heterostructure. (c) ARPES intensity map of the graphene-xenon Moir\'{e} heterostructure taken along the white-dashed line denoted in panel {\bf b}. The white arrow denotes the crossing point between two folded Dirac cones. (d) The second derivative of the ARPES spectrum shown in panel {\bf c}. Gray-dashed lines are guide to the eyes. (e-f) ARPES intensity maps of as-grown graphene (panel {\bf e}) and the graphene-xenon interface (panel {\bf f}) taken along the $k_{\rm x}$ direction denoted by an arrow with respect to the graphene unit cell shown in the inset. White curves are Lorentzian fits to the spectral intensity. (g) Energy-momentum dispersions of as-grown graphene (blue curve) and the graphene-xenon interface (red curve) shown as white curves in panels {\bf e} and {\bf f}. (h) Schematics of as-grown graphene (upper) and the graphene-xenon Moir\'{e} heterostructure (lower).}
  \label{fig:fig3}
\end{figure*}

Figure~3{\bf c} clearly shows folded Dirac cones induced by the graphene-xenon Moir\'{e} heterostructure in the ARPES intensity map taken along the $k_{\rm y}=0$ (white-dashed line in Fig.~3{\bf b}). Both Dirac cones are electron-doped by the substrate as discussed above. The crossing point between the Dirac cones that appears as a bright spot in the constant energy map taken at $E-E_{\rm F}=-1.0~{\rm eV}$ in Fig.~2{\bf a} is denoted by the white arrow in Fig.~3{\bf c}. Figure~3{\bf d} is the second derivative of the ARPES intensity map to enhance the low intensity features. The Dirac point of each cone does not show a clear signature of additional energy gap opening at the Dirac energy beyond what is expected from the as-grown graphene~\cite{JW_gap}, different from the lithium-adsorbed graphene case~\cite{Bao,Qu}.

The detailed analysis of the energy-momentum dispersion, however, exhibits a subtle but clear deviation from the Dirac dispersion of as-grown graphene. The crossing point of the Dirac cones exhibits a slight deformation as denoted by the white arrow in Fig.~3{\bf d}. This non-linearity may suggest that correlated electronic phases can be induced not only by the previous reported mechanically stacking of 2D crystals~\cite{Cao1,Cao2, Sharpe, Weston}, but also by producing Moir\'{e} potential simply with noble gases on 2D materials.

The energy-momentum dispersion taken along the $\Gamma$-K direction of the graphene unit cell gives information on electronic correlations. Figures~3{\bf e} and 3{\bf f} show ARPES intensity maps measured from as-grown graphene and the graphene-xenon heterostructure, respectively. Along this direction denoted by the arrow in the inset, only one of the two branches of the Dirac cone is observed due to the Berry phase of quasiparticles in graphene~\cite{HwangBerry}. The energy-momentum dispersion of each system was extracted from Lorentzian fit to momentum distribution curves (intensity spectrum taken at constant energy as a function of momentum) as shown in Fig.~3{\bf g}. The dispersion of as-grown graphene (blue curve) becomes steeper upon the formation of the graphene-xenon heterostructure (red curve) with increasing slope by about 8\% from 1.33$\times$10$^6$~m/s to 1.44$\times$10$^6$~m/s when extracted from the dispersion below the Dirac energy. Within the Fermi liquid theory, with which electron-doped graphene is well described~\cite{Sarma}, the enhanced slope indicates decreased effective mass of quasiparticles in graphene and hence suppressed electron correlations~\cite{Landau}. This is opposite from the behavior expected in the hybridized Dirac cones with strong electronic correlations in twisted graphene~\cite{Cao1,Cao2, Sharpe, Weston} or in the charge neutral Dirac cones where the Dirac energy lies at $E_{\rm F}$ ~\cite{HwangSR, HwangSTO}.

The suppressed electronic correlations can be understood via the enhanced dielectric screening from the xenon crystal formed on top of graphene. The 2D nature of graphene makes it very sensitive to its environment so the dielectrics efficiently tune electronic correlations~\cite{HwangSR,HwangSTO}. While the dielectric screening that as-grown graphene experiences is the average of that from SiC and vacuum, i.\,e.\,, $\epsilon_{\rm as-grown} = (\epsilon_{\rm vacuum}+\epsilon_{\rm SiC})/2$, graphene sandwiched by SiC and the xenon crystal will experience the dielectric screening as much as the average of that from SiC and xenon crystal, i.\,e.\,, $\epsilon_{\rm heterostructure} = (\epsilon_{\rm xenon}+\epsilon_{\rm SiC})/2$, as schematically shown in Fig.~3{\bf h}. Within the Fermi liquid theory, increased screening for Coulomb interactions will result in suppressed electronic correlations, so that the effective mass of quasiparticles decreases and hence the slope of an energy-momentum dispersion is expected to increase~\cite{Landau}. The dielectric constant of solid xenon has been reported to be $\epsilon_{\rm xenon}=1.37$~\cite{Steinberger}, which is larger than that of vacuum $\epsilon_{\rm vacuum}=1$. Naturally dielectric screening that graphene experiences will be stronger in the graphene-xenon heterostructure, resulting in the increased slope of the energy-momentum dispersion as witnessed in Fig.~3{\bf g}.









%

To further understand the electronic properties of the graphene-xenon heterostructure, the electronic band structure has investigated  by the first-principles calculations. As the first step, total energy of single xenon atom adsorbed on four graphene unit cells was calculated to find the most favorable adsorption site among the hollow, bridge, and on-top sites schematically shown in Fig.~4{\bf a}. Although the adsorption on the hollow site is still the most favorable in terms of energy, the energy differences are virtually negligible against the bridge and the on-top sites, with the values of $-$0.88~meV and $-$1.286~meV, respectively. As a result, thermal activation will promote the diffusion of xenon atoms on the graphene to form a 2D crystal, instead of favoring the adsorption of xenon atoms on randomly distributed hollow sites of graphene, especially when the formation energy of the xenon lattice ($\sim$12.0~meV) far exceeds the energy differences~\cite{Kern}.

\begin{figure*}[b]
  \includegraphics[width=1\columnwidth]{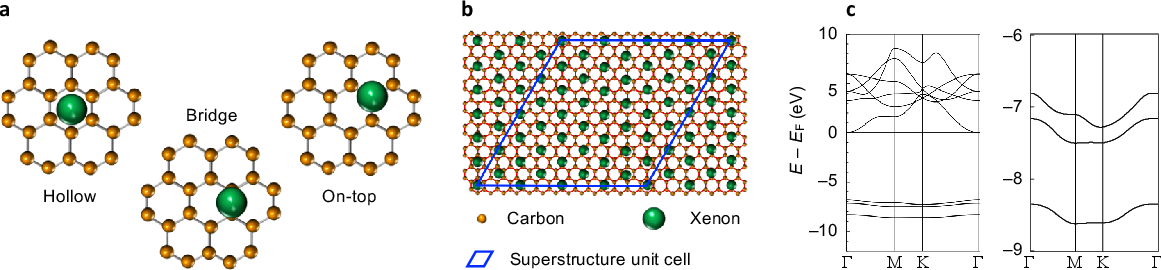}
  \caption{(a) Possible adsorption sites of a xenon atom on graphene. (b) The atomic configuration of the graphene-xenon interface, where orange and green balls denote carbon and xenon atoms, and the blue lozenge denotes the superstructure unit cell. (c) Calculated electron band structure of xenon over-layer depicted in panel (b) with a zoomed-in view at higher energy relative to the Fermi energy on the right panel.}
  \label{fig:fig4}
\end{figure*}

\medskip

Atomic configuration of the resultant graphene-xenon heterostructure is schematically shown in Fig.~4{\bf b}, where the blue lozenge denotes the unit cell of the Moir\'{e} heterostructure. It is easy to find that once one hollow site is occupied by a xenon atom, the next one sits slightly off the hollow site until the fifth one occupies the hollow site once again, constituting a superstructure with 13$\times$13 graphene unit cells and 5$\times$5 xenon unit cells. This is in excellent agreement with the observation of the well-defined LEED pattern shown in Fig.~1{\bf a} and folded Dirac cones around the $\Gamma$ point in ARPES measurements shown in Fig.~2{\bf a}. Figure~4{\bf c} shows the calculated electron band structure of the 2D xenon crystal formed on top of graphene. The most interesting thing observed from the calculated electronic band structure is that it does not show molecular states, i.\,e.\,, non-dispersive discrete states, that may be typically predicted from non-interacting gas particles. Instead, three bands, stemming from Xe 5{\it p} orbitals at an energy range from 7~eV to 8.5~eV below $E_{\rm F}$, have band width as broad as 0.7~eV as shown in the right panel, while rest of the bands above $E_{\rm F}$ are strongly dispersive. The dispersive electron bands support the formation of 2D xenon crystal as discussed with the LEED and ARPES results.

\medskip

\section{Conclusions}
The electronic structure of the graphene-xenon Moir\'{e} heterostructure has been investigated using ARPES and first-principles calculations. The formation of crystalline xenon has been evidenced by the additional LEED peaks and the folded six Dirac cones around the Brillouin zone center, which is further corroborated by the first-principles calculations. Upon the formation of the graphene-xenon hetero-interface, the slope of the Dirac cone increases due to the enhanced dielectric screening, i.\,e.\,, suppressed electronic correlations. Our findings suggest a viable route to create a Moir\'{e} heterostructure using the adsorption of noble gas on 2D materials, which goes beyond the conventional stacking and twisting of 2D layers. It also propose a possibility of tuning the electronic properties of 2D materials using the adsorbed layers of noble gases.

\acknowledgments
This work was supported by the National Research Foundation of Korea (NRF) grant funded by the Ministry of Science and ICT (No. 2021R1A2C1004266) and the National Research Facilities and Equipment Center (NFEC) grant funded by the Ministry of Education (No. 2021R1A6C101A429). The Advanced Light Source is supported by the Office of Basic Energy Sciences of the U.S. Department of Energy under Contract No. DE-AC02-05CH11231. K. K. was supported by KAERI internal R\&D program (524460-23). J.E.L. acknowledges support by Max Planck POSTECH/Korea Research Initiative, Study for Nano Scale Optomaterials and Complex Phase Materials (2022M3H4A1A04074153) through NRF funded by MSIP of Korea.

\end{document}